  \documentclass{gretsi}

\usepackage{bm}
 \usepackage{amsmath}
 \usepackage{amsfonts}
 \usepackage{microtype}
 \usepackage{url}
\usepackage[pdftex]{graphicx}
\usepackage[style=numeric,giveninits=true,backend=bibtex]{biblatex}
\bibliography{gretsifr.bib}

\usepackage[utf8]{inputenc}

\usepackage{lipsum}
\newcommand\blfootnote[1]{%
  \begingroup
  \renewcommand\thefootnote{}\footnote{#1}%
  \addtocounter{footnote}{-1}%
  \endgroup
}

 \usepackage[english,french]{babel}   
 \usepackage{times}			
 \usepackage{epsfig}
 \usepackage{csquotes}

\usepackage[T1]{fontenc}

\titre{Une ou deux composantes ? \\ La réponse de la diffusion en ondelettes}

\auteur{\coord{Vincent}{Lostanlen}{}}

\adresse{\affil{}{Music and Audio Research Lab, Université de New York\\
         35 West Fourth Street, New York, \'Etats-Unis}}

\email{vincent.lostanlen@nyu.edu}

\resumefrancais{Dans le contexte d'une modélisation biologiquement plausible de l'écoute artificielle, on s'intéresse à la représentation d'un signal stationnaire multicomposantes par un réseau de diffusion en ondelettes.
D'abord, nous montrons que la renormalisation des coefficients du second ordre par ceux du premier ordre donne un critère numérique simple pour établir si deux composantes de fréquences voisines interfèrent psychoacoustiquement.
Ensuite, nous généralisons le cadre théorique ``une ou deux composantes'' à trois sinusoïdes ou plus, et montrons en particulier qu'un réseau de diffusion en ondelettes de profondeur $M = \log_2 N$ suffit à caractériser l'amplitude relative des $N$ premiers termes d'une série de Fourier, tout en bénéficiant d'une propriété d'invariance à la transposition fréquentielle ainsi qu'au déphasage de chaque composante.}

\resumeanglais{With the aim of constructing a biologically plausible model of machine listening, we study the representation of a multicomponent stationary signal by a wavelet scattering network.
First, we show that renormalizing second-order nodes by their first-order parents gives a simple numerical criterion to assess whether two neighboring components will interfere psychoacoustically.
Secondly, we generalize the ``one or two components'' framework to three sine waves or more, and show that a network of depth $M = \log_2 N$ suffices to characterize the amplitudes of the first $N$ terms in a Fourier series, while enjoying properties of invariance to frequency transposition and component-wise phase shifts.}

\begin{document}
\maketitle

\section{Introduction}

En physiologie de l'audition, les cellules ciliées de notre cochlée jouent le rôle de filtres à facteur de qualité constant, dont la bande passante est large d'environ un quart d'octave.
Ainsi, étant données deux sinusoïdes $t \mapsto \bm{x_1}(t) = a_1 \cos(f_1 t + \varphi_1)$ et $t \mapsto \bm{x_2}(t) = a_2 \cos(f_2 t + \varphi_2)$ de fréquences respectives $f_1 > 0$ et $f_2 > 0$, nous percevons leur somme comme un accord de deux sons purs à condition que $\bm{x_1}$ et $\bm{x_2}$ appartiennent à des bandes critiques distinctes.
En revanche, si $a_2 \ll a_1$ ou $f_2 \approx f_1$, alors la composante $\bm{x_2}$ est dite \emph{masquée} par $\bm{x_1}$.
Au lieu de sons purs, nous entendons un ``battement'' : une onde localement sinusoïdale dont la fréquence porteuse est $\frac{1}{2} (f_1 + f_2)$ et la fréquence de modulation est $\frac{1}{2} \vert f_1 - f_2 \vert$.
La perception de cette modulation d'amplitude met en jeu des processus cognitifs ultérieurs à la cochlée, et notamment le cortex auditif primaire.

Or, la diffusion en ondelettes (\emph{wavelet scattering}) est un opérateur qui alterne, sur une échelle typique $T$, un banc de filtres passe-bandes analytiques à facteur de qualité constant $Q$, et l’application en tout point du module complexe.
Dès lors, ses deux premiers niveaux de profondeur modélisent respectivement les propriétés de la cochlée et du cortex auditif primaire.
Cet opérateur est employé à des fins de reconnaissance de la parole \cite{anden2014deep}, de sons environnementaux \cite{lostanlen2018jasmp}, de structures musicales répétées \cite{lostanlen2013atiam}, et de modes de jeux étendus \cite{lostanlen2018dlfm}.
La diffusion en ondelettes bénéficie donc, simultanément, d'une assise mathématique solide, d'applications innovantes en ingénierie informatique, et de liens privilégiés avec la neurophysiologie.

\blfootnote{Ce travail est financ\'{e} par la bourse ERC InvariantClass 320959.
Le code source reproduisant nos exp\'{e}riences et figures est en libre acc\`{e}s
a l'adresse \texttt{www.github.com/lostanlen/scattering.m}}

Cet article caractérise la réponse théorique de la diffusion en ondelettes au signal $\bm{x}(t) = \bm{x_1}(t) + \bm{x_2}(t)$.
À ce titre, il s'inscrit méthodologiquement dans une lignée de travaux antérieurs en traitement du signal non stationnaire, visant tous à étudier le comportement de tel ou tel opérateur convolutionnel non linéaire en fonction des grandeurs relatives $\frac{a_2}{a_1}$, $\frac{f_2}{f_1}$, et $(\varphi_2 - \varphi_1)$.
Trois de ces opérateurs, qui peuvent \^{e}tre mis en regard de la diffusion en ondelettes, sont la décomposition modale empirique \cite{rilling2008tsp}, le \emph{synchrosqueezing} \cite{wu2011aada}, et l'analyse spectrale singulière \cite{harmouche2015gretsi}.

Premièrement, nous établissons un critère numérique, appelé \emph{coefficient de masquage} $\boldsymbol{\kappa}$, fondé sur la diffusion en ondelettes d'ordre deux, et tendant vers $0$ pour $a_2 \rightarrow 0$ ou $\vert f_2 - f_1 \vert \rightarrow 0$ et proche de $1$ pour $a_1 \approx a_2$ et $0 < \vert f_2 - f_1 \vert \ll f_1$.
Ce coefficient de masquage présente, par construction, une invariance d'échelle : pour tout signal $\boldsymbol{x}$ et tous facteurs $\alpha$ et $\beta$ non nuls, il existe $T$ tel que $\boldsymbol{\kappa}(t \mapsto \alpha \boldsymbol{x}(\beta t)) \approx \boldsymbol{\kappa}(\boldsymbol{x})$.
Deuxi\`emement, nous g\'en\'eralisons le cadre théorique ``une ou deux composantes'' au cas de $N>2$ composantes $\bm{x_1}, \ldots, \bm{x_N}$, et \'etudions l'apport respectif de chaque ordre dans la diffusion en ondelettes de $\bm{x} = \sum_n \bm{x_n}$.
En particulier, nous démontrons qu'un réseau de profondeur $M = \log_2 N$ caractérise les $N$ premiers termes de la série de Fourier $\bm{x}(t) = \sum_n^N a_n \cos(n f_1 t + \varphi_n)$ par leurs amplitudes et fr\'equences respectives tout en \'etant invariant aux phases relatives $\varphi_n$.

\section{Interférométrie en ondelettes}

\begin{figure}
    \begin{center}
    \includegraphics[width=0.8\linewidth,keepaspectratio]{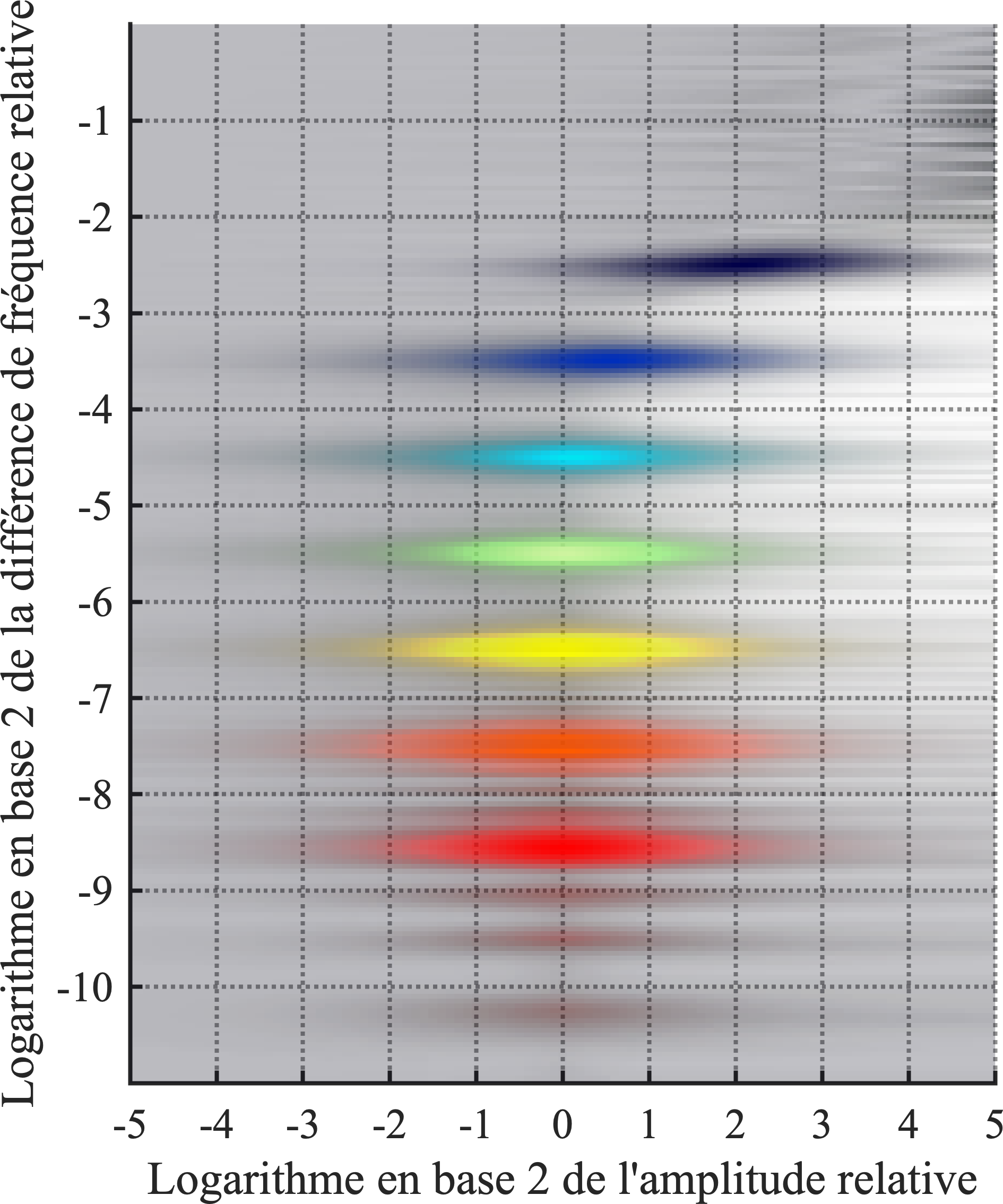}
    \end{center}
    \legende{Carte de chaleur du coefficient de masquage $\boldsymbol{\kappa}$ lors de la diffusion en ondelettes de deux composantes sinusoïdales $\bm{x_1}$ et $\bm{x_2}$, mesuré autour de la fréquence $f_1$, et évoluant en fonction de l'amplitude relative $\frac{a_2}{a_1}$ et de la différence de fréquence relative $\frac{\vert f_2 - f_1 \vert }{f_1}$. La couleur de chaque tache de diffusion indique la résolution d'analyse $\lambda_2$ de la deuxième strate du réseau de diffusion. Les ondelettes de diffusion ont un profil asymétrique (gammatone) et un facteur de qualité $Q=4$. Elles couvrent un intervalle total de neuf octaves en-deçà de $f_1$. Par souci de clarté, on affiche une interférence par octave au lieu de quatre. \label{fig:heatmap}}
\end{figure}

Soit $\bm{\psi}\in\mathbf{L}^2(\mathbb{R}, \mathbb{C})$ un filtre analytique de moyenne nulle, de fréquence centrale $1$, et de bande passante efficace $1/Q$. On définit un banc de filtres passe-bandes à facteur de qualité $Q$ constant comme la famille d'ondelettes $\bm{\psi_{\lambda}}: t \mapsto \lambda \bm{\psi} ( \lambda t)$
de fréquences centrales $\lambda$, de bandes passantes efficaces $\lambda/Q$, et d'\'echelles temporelles efficaces $2\pi Q/\lambda$.
En pratique, la variable fréquentielle $\lambda$ est discrétisée selon une série géométrique de raison $2^{1/Q}$.
Autrement dit, tout signal continu à bande limitée $[f_{\min}, f_{\max}]$ fait résonner au plus un nombre $Q \log_2({\frac{f_{\max}}{f_{\min}}})$ d'ondelettes $\boldsymbol{\psi_{\lambda}}$.
On définit le scalogramme comme le carré du module complexe de la transformée en ondelettes continue
\begin{align}
\mathbf{U_1}\bm{x}(t,\lambda_1) & =
\big \vert \bm{x} \ast \bm{\psi_{\lambda_1}} \big \vert^2 (t) \notag \\
& =
\left\vert \int_{\mathbb{R}} \bm{x}(t^{\prime}) \bm{\psi_{\lambda_1}}(t - t^{\prime})\;\mathrm{d}t^{\prime} \right \vert^2.
\end{align}
De même, on définit un second niveau de transformation non linéaire comme le ``scalogramme du scalogramme''
\begin{equation}
\mathbf{U_2} \bm{x} (t, \lambda_1, \lambda_2) =
\Big\vert \big \vert \bm{x} \ast \bm{\psi_{\lambda_1}} \big \vert^2 \ast \bm{\psi_{\lambda_2}} \Big \vert^2 (t).
\end{equation}
Une telle construction peut être itérée \emph{ad infinitum} en effectuant, pour tout entier naturel $m$, une ``diffusion'' (en anglais \emph{scattering}) du signal temporel multivarié $\mathbf{U_m}$ dans chacune des sous-bandes $\lambda_m < \lambda_{m-1}$ :
\begin{multline}
\mathbf{U}_{m+1} \bm{x}(t, \lambda_1 \ldots \lambda_{m+1}) = \\
\Big \vert \mathbf{U}_{m+1} \bm{x} (t, \lambda_1 \ldots \lambda_{m})
\ast \bm{\psi}_{\lambda_m} \Big \vert^2 (t, \lambda_1 \ldots \lambda_m).
\end{multline}
Puisque chaque strate $m$ du réseau s'exprime comme la composition d'un système linéaire invariant (la transformée en ondelettes continue) et d'une opération ponctuelle (le module complexe), on constate, par récurrence sur $m$, que tout tenseur $\mathbf{U}_m \bm{x}$ est équivariant à l'action du délai $\mathcal{T}_{\tau} : \bm{x} \mapsto \bm{x}(t - \tau)$.
Afin de remplacer cette propriété d'équivariance par une propriété d'invariance, on intègre $\mathbf{U_m}$ sur une durée pré-établie $T$, conduisant ainsi à la diffusion invariante (\emph{invariant scattering})
\begin{align}
    \mathbf{S}_{m} \bm{x}(t, \lambda) & =
    \int_{\mathbb{R}} \mathbf{U}_{m}(t^{\prime}, \lambda) \bm{\phi}_{T}(t-t^\prime)\;\mathrm{d}t^{\prime},
\end{align}
où le $m$-uplet $\lambda = (\lambda_1 \ldots \lambda_m)$ est un chemin de diffusion (\emph{scattering path}) et la fonction $\bm{\phi}_T$ un filtre passe-bas symétrique et réel de support temporel efficace $T$.


\section{Coefficient de masquage}

La convolution entre toute sinusoïde $\bm{x_n}$ et toute ondelette $\bm{\psi_{\lambda_1}}$ s'écrit comme une multiplication dans le domaine de Fourier. Puisque $\bm{\psi_{\lambda_1}}$ est analytique, seule la partie analytique $\bm{x_n^{\mathrm{a}}} = \bm{x_n} + \mathrm{i} \mathcal{H}\{\bm{x_n}\} = a_n \exp(\mathrm{i}(f_n t + \varphi_n))$ du signal réel $\bm{x_n}$ est préservée :
\begin{equation}
    \left(\bm{x_n} \ast \bm{\psi_{\lambda_1}}\right)(t)
    = \dfrac{1}{2} \bm{\widehat{\psi}_{\lambda_1}}(f_n) \bm{x_n^\mathrm{a}}(t).
\end{equation}
Par linéarité de la transformée en ondelettes continue, le cas à $N=2$ composantes s'exprime comme une multiplication hétérodyne faisant intervenir le conjugué complexe :
\begin{align}
    & \Big\vert ( \bm{x_1} + \bm{x_2} ) \ast \bm{\psi_{\lambda_1}} \Big\vert^2 (t) =
    \nonumber \\ & \mathfrak{R}\Bigg(\!\bm{\widehat{\psi}}\Big(\frac{f_1}{\lambda_1}\Big) \bm{\widehat{\psi}^\ast}\Big(\frac{f_2}{\lambda_1}\Big)\!\Bigg) a_1 a_2 \cos\big((f_2 - f_1) t + (\varphi_2 - \varphi_1) \big)
    \nonumber \\
    & + \frac{1}{2}  \Big\vert\bm{\widehat{\psi}}\big(\frac{f_1}{\lambda_1}\big) \Big\vert^2 a_1^2
    + \frac{1}{2}  \Big\vert\bm{\widehat{\psi}}\big(\frac{f_2}{\lambda_1}\big) \Big\vert^2 a_2^2.
    \label{eq:de-deux-composantes-a-une}
\end{align}

Puisque l'ondelette $\bm{\psi}$ est de moyenne nulle, les deux termes constants de l'équation ci-dessus, respectivement proportionels à $a_1^2$ et $a_2^2$, sont absorbés par le réseau à l'ordre un, et disparaissent à partir de la deuxième strate.
En revanche, le terme mixte, proportionnel à $a_1 a_2$, présente une fréquence fondamentale $\Delta\!f = \vert f_2 - f_1 \vert$.
Les auteurs d'une publication précédente ont remarqué que cette interférence produit un pic d'énergie au deuxième ordre de la diffusion pour $\lambda_1 = f_1$ et $\lambda_2 = \vert f_2 - f_1 \vert$ \cite{anden2012dafx}.
En revanche, ils ne commentent pas la dépendance de ce pic en terme de l'amplitude relative $\frac{a_2}{a_1}$, du profil de l'ondelette $\bm{\psi}$, du facteur de qualité $Q$, et de l'échelle temporelle de stationnarité locale $T$.
Cet article propose de remédier à ce manque, par une analyse plus exhaustive de ces différents paramètres.

Dans la définition originelle de la diffusion en ondelettes, la non-linéarité employée est l’amplitude complexe ($\vert z \vert = \sqrt{z \bar{z}}$) plutôt que la puissance ($\vert z \vert^2 = z \bar{z}$) , afin de garantir que chaque opérateur soit Lipschitz-contractant \cite{mallat2012cpam}.
Néanmoins, pour simplifier les calculs et épargner une étape de linéarisation de la racine carrée, nous choisissons d'adopter une mesure de puissance plutôt que d’amplitude.
Cette idée a été initialement proposée par \cite{balestriero2017arxiv} dans un contexte de bioacoustique sous-marine.

L'équation \ref{eq:de-deux-composantes-a-une} montre que le premier ordre de diffusion transforme le signal à deux composantes $\boldsymbol{x} = \boldsymbol{x_1} + \boldsymbol{x_2}$ en un signal à une composante.
Pour que cette composante soit non négligeable, trois conditions doivent être réunies.
Premièrement, le produit $a_1 a_2$ doit être non négligeable devant chacun des carrés $a_1^2$ et $a_2^2$.
Deuxièmenent, il doit exister une résolution d'analyse $\lambda_1$ contenant les deux fréquences $f_1$ et $f_2$ dans sa bande passante.
Autrement dit, $\lambda_1$ doit satisfaire l'inégalité $\vert \frac{f_n}{\lambda_1} - 1 \vert < \frac{1}{Q}$ pour $f_n = f_1$ et $f_n = f_2$ à la fois.
Troisièmement, la différence de fréquence $\Delta\!f$ doit appartenir à la bande passante d'une ondelette $\bm{\psi_{\lambda_2}}$ au sein de la deuxième strate du réseau de diffusion.
Or, en pratique, afin d'assurer la localisation temporelle des coefficients et de restreindre chaque banc de filtres à un nombre fini d'octaves, la dilatation des ondelettes $\bm{\psi_{\lambda_m}}$ est bornée par la constante temporelle $T$.
Par conséquent, il est nécessaire que la période $\frac{2\pi}{\vert f_2 - f_1\vert}$ du signal différentiel soit inférieure à la pseudo-période de l'ondelette de support efficace $T$, c'est-à-dire $Q T$. On a donc nécessairement $\vert f_2 - f_1 \vert < \frac{2 \pi Q}{T}$.

Une manière simple de mesurer le degré d'interférence mutuelle des composantes $\boldsymbol{x_1}$ et $\boldsymbol{x_2}$ consiste à renormaliser les coefficients du second ordre par ceux du premier ordre:
\begin{equation}
    \mathbf{\widetilde{S}_2}\boldsymbol{x}(t,\lambda_1,\lambda_2) = \dfrac{\mathbf{S_2}\boldsymbol{x}(t, \lambda_1, \lambda_2)}{\mathbf{S_1}\boldsymbol{x}(t, \lambda_1)}
\end{equation}
Une telle opération de post-traitement, proposée par \cite{anden2014deep}, se rapproche conceptuellement de méthodes classiques de contrôle adaptatif du gain, et notamment de la normalisation d'énergie par canal (\emph{per-channel energy normalization} ou PCEN) \cite{lostanlen2019spl}.

En accord avec la méthodologie ``une ou deux composantes'' \cite{rilling2008tsp}, la figure \ref{fig:heatmap} illustre la valeur de ce rapport d'énergie dans la bande de fréquence $\lambda_1 = f_1$, pour différentes valeurs de l'amplitude relative $\frac{a_2}{a_1}$ et de la différence de fréquence relative $\frac{\vert f_2 - f_1 \vert}{f_1}$.
On a fixé $f_2 < f_1$ sans perte de généralité.
Comme attendu, on constate que, pour $a_2 \approx a_1$ et une différence de fréquence relative comprise entre $\frac{Q}{f_1 T}$ et $\frac{1}{Q}$, les ondelettes de la seconde strate $\bm{\psi_{\lambda_2}}$ résonnent avec le signal différentiel produit par l'interférence des composantes $\bm{x_1}$ et $\bm{x_2}$.
La différence des logarithmes entre $\lambda_1$ et $\lambda_2$ caractérise l'intervalle musical formé par ces deux composantes.
De plus, afin d'approximer la réponse des cellules ciliées de la cochlée, on emploie des ondelettes Gammatone \cite{lostanlen2018jasmp}.
Ce choix a pour conséquence de produire un masquage fréquentiel asymétrique, : puisque $f_2 < f_1$ par convention, il est possible, à $a_1$ et $a_2$ fixées, que $\boldsymbol{x_1}$ soit dans la bande critique de $\boldsymbol{x_2}$ sans que l'inverse soit nécessairement vrai.
L'asymétrie est d'autant plus prononcée que la différence de fréquence est grande.
Ce phénomène est en accord avec des résultats expérimentaux connus en psychoacoustique.

Afin de décider si la composante $\boldsymbol{x_2}$ est masquée par $\boldsymbol{x_1}$, on définit le coefficient de masquage
\begin{equation}
    \boldsymbol{\kappa}(\boldsymbol{x}) : t \mapsto \sum_{\lambda} \mathbf{\widetilde{S}_2}\boldsymbol{x}(t,\lambda_1,\lambda_2)
\end{equation}
comme la somme des coefficients renormalisés à travers tous les chemins de diffusion en ondelettes $\lambda = (\lambda_1, \lambda_2)$.
Pour $T$ assez grand, le critère $\boldsymbol{\kappa}$ bénéficie de trois propriétés d'invariance : (i) au déphasage $\mathcal{P}_{\theta_1, \theta_2} : (\varphi_1, \varphi_2) \mapsto (\varphi_1 + \theta_1, \varphi_2 + \theta_2)$; (ii) à la transposition fréquentielle $\mathcal{F}_\gamma : (f_1, f_2) \mapsto (2^{\gamma} f_1, 2^{\gamma} f_2)$ ; et (iii) au changement d'intensité $\mathcal{I}_{\gamma} : (a_1, a_2) \mapsto (2^{\gamma} a_1, 2^{\gamma} a_2)$.

\section{Trois composantes ou plus}

\begin{figure}[htb]
    \begin{center}
    \epsfig{file=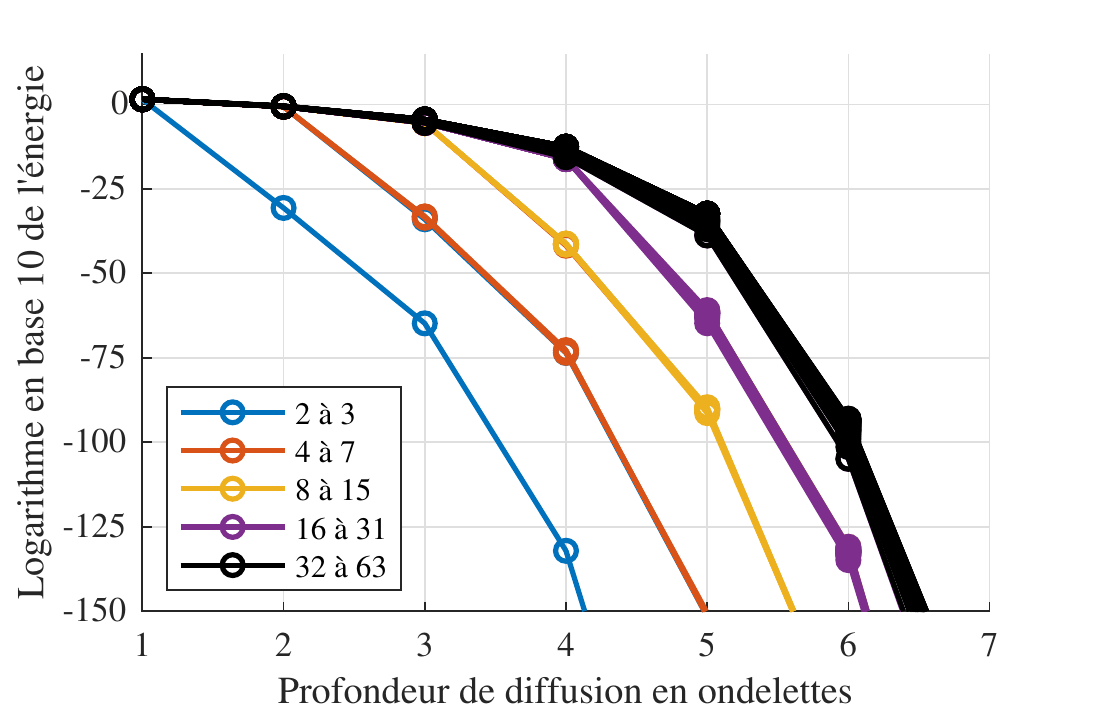,width=88mm}
    \end{center}
    \legende{Courbes de décroissance de l'énergie diffusée en fonction de la profondeur $m$ du réseau de diffusion en ondelettes, pour des signaux à $n$ composantes d'amplitude égale et de fréquences arithmétiquement espacées. La couleur de chaque courbe indique la partie entière de $\log_2 n$. Les ondelettes de diffusion ont un profil en sinus cardinal (ondelettes de Shannon) et un facteur de qualité $Q=1$. À chaque niveau de profondeur, le banc de filtres couvre un intervalle total de sept octaves.}
\end{figure}

En acoustique musicale, les sons tonaux naturels sont rarement approximables comme un mélange de seulement deux composantes. Bien plus souvent, ils comportent dix composantes, voire plus, et couvrent simultanément plusieurs octaves.
Dès lors, un calcul à l’ordre deux des coefficients de masquage n’apporte qu’une représentation grossière du contenu timbral de chaque bande critique. En effet, ces coefficient encodent les relations mutuelles entre paires de composantes, mais demeurent peu sensibles à la présence de structures plus globales dans l’enveloppe spectrale de la note de musique analysée.

Dans un contexte d’écoute artificielle, l’identification d’une source peut se formuler --- en régime stationnaire, du moins --- comme un problème inverse sur les paramètres physiques de l’équation des ondes, sous contrainte d’invariance à la fréquence fondamentale.
S’agissant d’instruments de musique, on sait par exemple que le mode d’excitation (corde pincée ou frappée) influence l’exposant de décroissance de l’amplitude en fonction de la fréquence, tandis que les conditions aux bords (conduit fermé ou semi-ouvert) influence l’amplitude relative des harmoniques paires par rapport aux harmoniques impaires.

La construction d’un espace de similarité dans lequel les deux facteurs de variabilité susnommés se traduisent par des effets quasi linéaires, orthogonaux, invariants à la fréquence fondamentale, et robustes à la présence de bruit, est essentielle à l'amélioration des performances en apprentissage automatique des sons musicaux.
Pourtant, la difficulté de sa résolution réside dans son aspect multi-échelles.
Si l’on restreint l’interférométrie à deux partiels adjacents en fréquence, il est difficile de distinguer les effets respectifs des conditions initiales (onde de forme rectangulaire ou triangulaire) et des conditions aux bords (onde à symétrie paire ou impaire) sur les paires d’amplitudes observées.
Dans le but de caractériser l’effet conjoint de paramètres physiques divers sur la production d’une note de musique, nous proposons d’étendre la procédure de diffusion en ondelettes au-delà de l’ordre deux. À partir de l’ordre trois, les interférences ne mettent plus en jeu des paires de composantes, mais des structures hétérodynes de plus grande multiplicité, quoique restreintes à une bande critique de largeur préétablie $1/Q$, mesurée en fraction d’octave.

L’application alternée d’une transformée en ondelettes continue et du module complexe présente une forte ressemblance avec l’architecture des réseaux de neurones convolutifs. On parle donc de réseau de diffusion (\emph{scattering network}), organisé selon différents niveaux (\emph{layers}) de profondeur. Intuitivement, le passage d’un niveau au suivant double la multiplicité de l’interférence, exprimée en termes de nombre de composantes.
La figure 2 illustre cette dépendance logarithmique entre le nombre de composantes et la profondeur de diffusion maximale.
On se place dans le cas particulier de la série de Fourier $\bm{x}: t \mapsto \sum_n^N \cos(n f_1 t)$, c'est-à-dire $a_n = a_1$, $f_n = n f_1$, et $\varphi_n = 0$ pour tout entier $n$.
On étudie la norme $\ell^2$ de chaque strate, obtenue en sommant les chemins $\lambda = (\lambda_1 \ldots \lambda_M)$ de profondeur finie $M$.
Pour $N$ compris entre $2^M$ et $2^{M-1}$, on constate que les $M$ premières strates du réseau de diffusion parviennent à absorber la quasi-totalité de l'énergie totale du signal, l'énergie résiduelle tombant à zéro pour $M > \log_2 N$.

Pour justifier cette observation, on raisonne par récurrence sur la profondeur du réseau de diffusion. On décompose la partie analytique de $\bm{x}$ en une somme de termes $\bm{y_J}$ dont la bande passante débute à la fréquence $2^J$ et s'arrête à la fréquence $2^{J+1}$.
L'utilisation d'ondelettes analytiques de Shannon (profil symétrique en sinus cardinal) avec $Q=1$ permet de séparer les signaux $\bm{y_J}$, chacun comprenant $2^J$ composantes.
L'application du module carré s'écrivant comme une convolution dans le domaine de Fourier, $\vert \bm{y_J} \vert^2$ est une série de Fourier de fréquence fondamentale $\frac{f_1}{2}$ et de bande passante $[-2^{J}; 2^{J}]$.
On se ramène ainsi à l'hypothèse de récurrence avec une profondeur $(M-1)$ et au plus $\frac{N}{2}$ composantes, ce qui permet de conclure.


\section{Conclusion}

Dans cet article, on étudié le rôle de chaque strate dans un réseau de diffusion à partir d'une méthodologie de type ``une ou deux composantes'' \cite{rilling2008tsp}.
On a donné un critère numérique de masquage psychoacoustique fondé sur la diffusion en ondelettes de sons différentiels.
Dans le cas d'une série de Fourier, on a montré que l'énergie diffusée tombe subitement à zéro à partir d'une profondeur de l'ordre du logarithme du nombre de composantes.
Ce résultat est un cas particulier du théorème de décroissance exponentielle des coefficients de diffusion \cite{waldspurger2017exponential}, qui est généralement valable dans $\mathbf{L}^2$ mais exprimé en terme de borne supérieure d'énergie à profondeur fixée, non de borne supérieure de profondeur à bande passante fixée.

\printbibliography

\end{document}